\documentclass[preprint,3p,times,authoryear]{elsarticle}

\usepackage{graphicx}
\usepackage{amsmath}
\usepackage{amssymb}
\usepackage{amsfonts}
\usepackage{mathrsfs}
\usepackage{adjustbox}
\newcommand{\q}[1]{``#1''}
\usepackage{mdframed}
\usepackage[figuresright]{rotating}
\usepackage{courier} 
\usepackage[final]{changes}

\usepackage{setspace}

\usepackage{dutchcal}

\graphicspath{ {figures/} }

\usepackage{lineno}

\usepackage[utf8x]{inputenc} 

\usepackage{nomencl}
\usepackage{etoolbox}
\usepackage{wrapfig}
\usepackage{stfloats} 
\usepackage{framed} 
\usepackage[skins,breakable, most]{tcolorbox}

\renewcommand*\nompreamble{\begin{multicols}{2}} 
\usepackage{epstopdf} 
\usepackage{todonotes}
\usepackage{setspace}
\usepackage{siunitx}

\usepackage[caption = false]{subfig}
\usepackage[hyphens]{url}
\usepackage{graphicx}
\usepackage{booktabs}

\usepackage{lmodern,textcomp}
\usepackage{rotating} 
\usepackage{multicol} 
\usepackage{multirow} 
\usepackage{array} 
\usepackage{mathtools, cuted}
\usepackage{lscape}
\usepackage{tabularx} 
\usepackage{longtable}  
\usepackage[flushleft]{threeparttable}
\usepackage{threeparttablex} 
\usepackage{tabulary}
\usepackage{amssymb}
\usepackage{pifont}
\RequirePackage[ngerman=ngerman-x-latest]{hyphsubst}
\usepackage{microtype}
\usepackage[american]{babel} 
\usepackage{hyperref}
\usepackage{float}
\usepackage[normalem]{ulem}
\useunder{\uline}{\ul}{}

\usepackage{natbib}
\biboptions{square,comma,sort&compress}

\setcitestyle{numbers}
\journal{}
\makeatletter
\def\ps@pprintTitle{%
 \let\@oddhead\@empty
 \let\@evenhead\@empty
 \def\@oddfoot{}%
 \let\@evenfoot\@oddfoot}
\makeatother

\usepackage[acronym]{glossaries}
\usepackage{glossary-inline}
\renewcommand*\nompreamble{\begin{multicols}{2}}
\makeglossaries
\newacronym{LCT}{LCT}{low-carbon technology}
\newacronym{GHG}{GHG}{greenhouse gas}
\newacronym{TPB}{TPB}{Theory of Planned Behavior}
\newacronym{DOI}{DOI}{Diffusion of Innovation Theory}
\newacronym{VBN}{VBN}{Value-Belief-Norm Theory}
\newacronym{PBC}{PBC}{perceived behavioral control}
\newacronym{PV}{PV}{photovoltaic system}
\newacronym{EEA}{EEA}{energy efficient appliance}
\newacronym{GT}{GT}{green tariff}
\newacronym{EV}{EV}{electric vehicle}
\newacronym{WTP}{WTP}{willingness to pay}
\newacronym{TCO}{TCO}{total cost of ownership}

\newcommand\blfootnote[1]{%
 \begingroup
 \renewcommand\thefootnote{}\footnote{#1}%
 \addtocounter{footnote}{-1}%
 \endgroup
}

\begin{document}
\begin{frontmatter}


\title{Product traits, decision-makers, and household low-carbon technology adoptions: moving beyond single empirical studies}


\author[IIRM]{Emily Schulte\corref{cor1}}
\ead{schulte@wifa.uni-leipzig.de}
\author[IIRM,DTU]{Fabian Scheller}
\author[UF]{Wilmer Pasut}
\author[IIRM]{Thomas Bruckner}
\cortext[cor1]{Corresponding author}


\address[IIRM]{Institute for Infrastructure and Resources Management (IIRM), Leipzig University}
\address[DTU]{Energy Economics and System Analysis, Division of Sustainability, Department of Technology, Management and Economics, Technical University of Denmark (DTU)}
\address[UF]{Department of Environmental Sciences, Informatics and Statistics, Universita Ca' Foscari}

\begin{abstract}
Although single empirical studies provide important insights into who adopts a specific LCT for what reason, fundamental questions concerning the relations between decision subject (= who decides), decision object (= what is decided upon) and context (= when and where it is decided) remain unanswered.
In this paper, this research gap is addressed by deriving a decision framework for residential decision-making, suggesting that traits of decision subject and object are determinants of financial, environmental, symbolic, normative, effort and technical considerations preceding adoption. Thereafter, the decision framework is initially verified by employing literature on the adoption of photovoltaic systems, energy efficient appliances and green tariffs. Of the six proposed relations, two could be confirmed (financial and environmental), one could be rejected (effort), and three could neither be confirmed nor rejected due to lacking evidence.
Future research on LCT adoption could use the decision framework as guidepost to establish a more coordinated and integrated approach, ultimately allowing to address fundamental questions.
\end{abstract}



\end{frontmatter}

\section*{Highlights}
\begin{itemize}
    \item Decision subject and object are equal parts of residential adoption behavior
    \item Decision framework suggests that traits of decision subject and object predict the evaluation of the object
    \item Proposed relations for financial and environmental considerations confirmed
    \item Blind spots concerning normative and technical considerations
    \item Policies should target decision subject and object to enhance adoption rates
\end{itemize}

\blfootnote{\printglossary[type=\acronymtype,style=inline,nonumberlist]}


\section{Introduction}
\label{S:1}

\subsection{Motivation and problem statement}
\label{S:1.1}
Residential \acrfull{LCT} uptake, subsuming the household level adoption of "innovative consumer-side technologies [...] that have the potential to reduce residential \acrfull{GHG} emissions by decreasing overall demand for carbon-intensive energy from fossil fuels" \cite{Scheller.2020}, is important to reaching the aspired global zero emission economy by 2050 \cite{Geels.2018, Frederiks.2015}. Their uptake depends on residential decision-making, and a large amount of studies dealing with the motivations and barriers to adopt single LCTs exists. Researchers employ different methods, theoretical backgrounds and explanatory variables to explain interest, purchase intention, willingness to pay or actual behavior, and conduct studies at different times in different places. Although single studies provide important insights into who adopts a specific LCT for what reason, fundamental questions still remain unanswered. For example, why does the broad majority of consumers doesn't adopt LCTs that are economically feasible in the long term, e.g., photovoltaic systems (\acrshort{PV}) and electric vehicles (\acrshort{EV}), and, on the other hand, why do consumers adopt a product that has negative financial consequences, e.g., a green electricity tariff (\acrshort{GT})? Is the adoption of such LCTs primarily driven by environmental considerations? To what extent is the adoption of conspicuous LCTs driven by status aspirations, and how is the lack of visibility compensated in the case of LCTs that are consumed in private? How does the diffusion stage of an LCT influences adoption? Do the initial and expected long-term effort related to LCTs hinder adoption?

All these questions that could be raised to fundamentally understand residential LCT adoption decisions revolve around the relations between decision subject (= who decides), decision object (= what is decided upon) and context (= when and where it is decided). Yet, current studies focus predominantly on the decision subject and its subjective perception of the decision object, whilst traits of the decision object \cite{Fleiter.2012, Faiers.2007b} and the context in which the decision is embedded \cite{Geels.2018, Wilson.2007, Axsen.2012} have not received systematic attention. Thus, no basic structure describing relations between subject, object and context within the process of residential decision-making exists \cite{Faiers.2007b}.

\subsection{Subject, object and context in residential LCT decision-making}
\label{S:1.2}
Typical studies on LCT decision-making are focused on the decision subject, and its subjective perception of the object. More precisely, they use socio-demographic and psychographic characteristics, previous behavior, and the respondents' perception or evaluation of the LCT to explain intention to adopt or actual behavior (see e.g. \cite{Schleich.2019, Ziegler.2020, Alipour.2020}). This focus can partly be explained by the employed theoretical backgrounds. Oftentimes, the framework of the \acrfull{TPB} \cite{Ajzen.1991} is used and extended, and ideas of the \acrfull{DOI} \cite{Rogers.2003} and the \acrfull{VBN} \cite{Stern.2000}, accounting for innovative and pro-environmental aspects of the products respectively, are integrated.

DOI and VBN offer different characteristics of the decision subject as explanatory variables. DOI suggests that socio-demographic (e.g. education, income) and psychographic (e.g. innovativeness, cosmopoliteness) measures vary between first adopters, called innovators, and laggards who adopt an innovation last \cite{Rogers.2003}, whereas VBN proposes to explain and predict pro-environmental behavior using characteristics of individuals related to their attitude towards the environment \cite{Stern.2000}. However, no prototypical green consumer exists \cite{Peattie.2010}, socio-demographics are no good predictors of specific behaviors \cite{Ajzen.1991, Arts.2011, Hansla.2008}, and pro-environmental attitudes are not necessarily translated into behavior \cite{Peattie.2010, Curtis.2018, Trotta.2018}, which calls into question the usefulness of individual characteristics as direct predictors of LCT adoption.

In addition, DOI and TPB suggest that the decision subjects subjective perception of an LCT is helpful to explain behavior. In DOI, the perception of the innovations relative advantage, compatibility, complexity, observability and trialability are central predictors for the speed of diffusion in a society, whereas intrinsic attributes of the object are secondary \cite{Rogers.2003}. In TPB, the prediction of intention and behavior relies on the constructs attitude, subjective norm and \acrfull{PBC}, each representing the result of unspecified evaluative criteria decision subjects apply to the object \cite{Ajzen.1991}. Yet, why do decision subjects perceive objects differently? According to Axsen and Kurani \cite{Axsen.2012}, TPB \q{offers little explanation of the origins and dynamics of these individuals’ attitudes and [...] social norms}, and the same applies to DOI. Moreover, perceived variables lose their explanatory power beyond the particular study framing, as decision subjects evolve, change behavior and their living situation, decision objects may become cheaper and more common, and political, social and infrastructural contexts change over time \cite{Ajzen.1991, Rogers.2003, Jansson.2011}.\\

On the other hand, decision-making has been described as \q{a goal-directed process in which consumers evaluate [...] attributes with certain use purposes and situations in mind} \cite{Arts.2011}, and Wolske et al.\cite{Wolske.2017} suggest that the evaluation of an LCT is the result of a subjective assessment of its advantages and disadvantages in which the decision-makers \q{broad dispositions may influence how an innovation is perceived}. Both thus suggest that a decision subject evaluates an object before adoption, which appears to be a simple, but realistic description of decision-making. It accounts for differences among decision subjects - the same object can be evaluated differently, and it accounts for differences among decision objects - the same subject can evaluate objects differently.
Moreover, explicitly including the decision object is a mean to overcome the criticism of a number of researchers that individual behavior is oftentimes analyzed in an empty space \cite{Geels.2018, Wilson.2007, Axsen.2012, Wolske.2020}, as it relates the decision to a context. For example, diffusion stages vary across places and change over time \cite{Baginski.2019}, increasing the attractiveness of an LCT for individuals who are more like laggards and less like innovators \cite{Rogers.2003}, and investment costs or payback periods can be affected by policy measures, influencing how an individual evaluates the financial outcomes of adoption.

\begin{figure}[ht]
    \centering
    \includegraphics[width=0.75\textwidth]{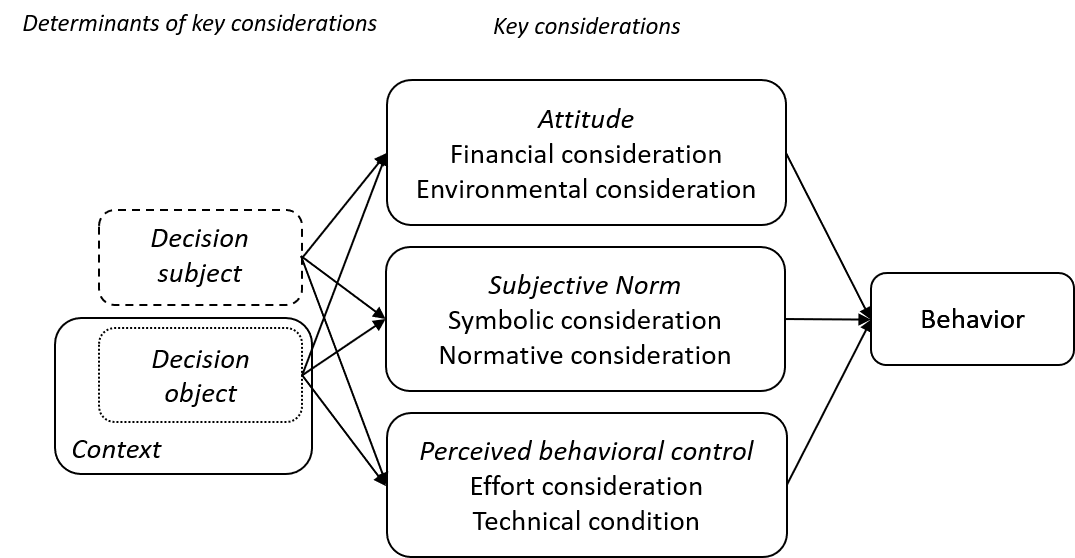}
    \caption{In the general framework, decision subject and decision object are predictors for the subjects perception of the object, which in turn predicts behavior. Traits of the decision object reflect its embedding in a context.}
    \label{fig:gen_struc}
\end{figure}

Following Arts et al. \cite{Arts.2011} and Wolske et al. \cite{Wolske.2017}, we suggest a basic structure to explain residential decision-making, working towards understanding underlying relations between decision subject, decision object and context. The structure is depicted in Figure \ref{fig:gen_struc}. Decision subject and object are predictors for the subjects perception of the object, which in turn predicts behavior. The perception is conceptualized based on the TPB constructs attitude, subjective norm and PBC. To overcome the problem that the proposed constructs do not provide information about the concrete considerations of the decision subject, they are further specified in six central, thematically distinct considerations: The attitude towards an LCT is likely to be affected by an evaluation of financial and environmental outcomes of behavior \cite{Palm.2018, Rezvani.2015}. Potential gains in status or identity, and the timing of adoption relative to one's peers relate to subjective norms \cite{Axsen.2012, Wolske.2020}, and practical considerations related to the effort required to adopt and use an LCT, and the fulfillment of local requirements for the LCT are central evaluative criteria that can control behavior \cite{Palm.2018, Galvin.2020}. According to Axsen and Kurani \cite{Axsen.2012}, considerations can outweigh each other, as for example, a perceived environmental benefit or increased status might outweigh additional financial costs.

\subsection{Research objective and structure}
\label{S:1.3}
This research paper is an initial effort towards a fundamental understanding of relations between the decision subject, decision object and context in the context of residential LCT adoption. We pursue two main objectives: 

Firstly, we derive a decision framework to explain residential decision-making, taking into account subject, object and context (Section \ref{S:3}). For this purpose, we further develop the basic structure already derived (Figure \ref{fig:gen_struc}), and add traits of decision subject and object that might play a role in the six central considerations. The traits are derived within a directed content analysis.

Secondly, we use three example LCTs (PV, EEA, GT) to examine the extent to which the proposed decision framework can already be confirmed or rejected by the existing literature on actual adoption behavior (Section \ref{S:4}). To this end, the traits of the example LCTs are evaluated (Section \ref{S:4.1}). Then, we systematically examine which traits of the decision subject and which considerations are used in the literature (Section \ref{S:4.2}), and which effects were found (Section \ref{S:4.3}). Based on the results of the evaluation of the example LCTs and the systematic literature analysis, we then discuss to what extent the proposed decision framework can be confirmed or rejected (Section \ref{S:5.1}).

Thereafter, we discuss shortcomings (Section \ref{S:5.2}) and elaborate on the merits of a coordinated and integrated approach in LCT adoption research (Section \ref{S:5.3}). The paper closes with a short conclusion (Section \ref{S:6}).\\

The conducted research contributes to the scientific debate in three ways. First, the derived decision framework draws together different theoretical approaches and the variety of explanatory variables employed in LCT adoption literature, providing a fundamental and practical framework for understanding residential LCT adoption decisions. It goes beyond the work of Faiers et al. \cite{Faiers.2007b} and Alipour et al. \cite{Alipour.2020}, as it combines theoretical approaches on the one hand, and on the other hand takes up the wealth of explanatory variables. Second, the analysis illuminates blind spots of the literature body on residential LCT adoption, and shows that a systematic integration of decision subject, object and context is useful to generate insights in residential LCT decision-making beyond the results of single empirical studies. Third, the decision framework can serve as guidepost for future studies. Depending on the locally specific traits of the object under consideration, the decision framework can be used to hypothesize relationships between relevant variables. The resulting more integrated and coordinated study design across LCTs could increase the overall comparability of studies, working towards comprehensive understanding of residential LCT adoption.
\section{Methods}
\label{S:2}

\subsection{Derivation of the decision framework}
\label{S:2.1}

In order to reach the first objective, we develop the basic structure (Figure \ref{fig:gen_struc}) into a decision framework.

To identify and define relevant traits of decision subject and object relating to the six key considerations (financial, environmental, symbolic, normative, effort and technical issues), the literature body on residential LCT adoption is reviewed (Section \ref{S:3}). Papers assessing the explanatory power of predictors for various dependent variables (e.g., awareness, interest, intention, attitudes, motives, willingness, adoption behavior) and the variety of LCTs are included in this analytical step, as also done by Selvakkumaran and Ahlgren \cite{Selvakkumaran.2019} and Alipour et al. \cite{Alipour.2020}. In an ordinary search of peer-reviewed scientific literature, keywords related to LCTs are combined with keywords related to residential decision-making (adoption decision, adoption, household transition, driver, barrier, explanatory variables, behavioral theory). Building on the procedure of Alipour et al. \cite{Alipour.2020} and Selvakkumaran and Ahlgren \cite{Selvakkumaran.2019}, the methodology of a directed content analysis \cite{Hsieh.2005} is followed. A decision framework with seven subject and nine object traits is proposed in Figure \ref{fig:ext_struc}. Moreover, we propose explicit relations between the expression of traits in subject and object and the considerations.

\subsection{The decision framework in current quantitative studies}
\label{S:2.2}

In the second stage of the analysis, we examine the extent to which the proposed decision framework can already be confirmed or rejected by the existing literature on actual adoption behavior.

For the systematic analysis, photovoltaic systems (PV), energy efficient appliances \acrshort{EEA} and GT are selected as exemplary, very different LCTs. Only such papers focusing on behavior will be included to avoid biases due to the intention-behavior gap \cite{Arts.2011, Peattie.2010, Carrington.2014}.
An ordinary search of quantitative, peer-reviewed scientific literature researching adoption decisions concerning PV (residential solar photovoltaic system, residential photovoltaic, residential PV, domestic solar photovoltaic system, domestic photovoltaic, domestic PV, microgeneration), EEA (energy efficient appliance, efficient home appliance) and GT (green tariffs, green electricity, green energy) was performed. 19 papers have been identified; 8 on PV, 8 on EEA and 3 on GT.

Based on explicit and indirect statements in the identified literature, we evaluate the example LCTs concerning the expression of the nine traits of the decision object proposed in the decision framework (Section \ref{S:4.1}). Table \ref{tab:lct_evalu} summarizes the evaluation.
Then, we systematically examine which traits of the decision subject and which considerations are used in the literature (Section \ref{S:4.2}), and which effects were found (Section \ref{S:4.3}).
In an Excel database, explanatory variables used in the identified papers are clustered into the respective key consideration (Table \ref{tab:key_cons}) or trait of the decision subject (Table \ref{tab:indi_chara}). A simple counting methodology is used to reveal what aspects of the decision framework are addressed. Moreover, five different values are assigned to summarize findings for the covered aspects, broadly in line with Alipour et al. \cite{Alipour.2020}. If an explanatory variable is assessed with backward coding (e.g. lack of environmental concern), the relationship is reversed to maintain consistency. The evaluation methodology is summarized in Table \ref{tab:vari_evalu}. The clustered original variables and assigned values are presented in Tables \ref{tab:key_cons} and \ref{tab:indi_chara}, incorporating key considerations and traits of the decision subject respectively.

\begin{table}[htbp]
    \footnotesize
    \centering
  \caption{Coding scheme and description}
    \begin{tabular}{p{5em}p{35em}}
    \toprule
    \textbf{Code} & \textbf{Description}  \\
    \midrule
    $++$ & Significant positive relationships: Positive $\beta$ value with p-values $\leq$ 0.05 \\
    \midrule
    $+$ & Insignificant positive relationships: Driving force \\
    \midrule
     $o$ & No positive or negative relationships: No positive or negative relationship \\
    \midrule
    $-$ & Insignificant negative relationships: Barrier  \\
    \midrule
    $--$ & Significant negative relationships: Negative $\beta$ value with p-values $\leq$ 0.05  \\
    \bottomrule
    \end{tabular}
  \label{tab:vari_evalu}
\end{table}
\section{A decision framework for LCT adoption behavior}
\label{S:3}

\subsection{Financial considerations}
\label{S:3.1}
When considering an LCT for adoption, the financial effects of the behavior on the households financial means are likely to be assessed. High investment costs have been found to hinder the adoption of PV systems \cite{Jacksohn.2019}, heating systems \cite{Curtis.2018} and electric vehicles \cite{Danielis.2018, Rezvani.2015}, and the extra costs for enabling technologies are negatively correlated with the \acrfull{WTP} for green energy and dynamic tariffs \cite{KowalskaPyzalska.2018b}. In this context, Mills and Schleich \cite{Mills.2012} propose to distinguish between \q{low-cost or no-cost measures [...] and high-cost measures which require capital investment [...]}. Despite one could suspect that LCTs with high upfront costs are adopted particularly from high-income consumers, results in literature are heterogeneous: Whereas no difference in income could be found between adopters and non-adopters of alternative fuel vehicles in Sweden \cite{Jansson.2011}, and likely and not-likely adopters of PV systems in Switzerland \cite{Petrovich.2019}, Jacksohn et al. \cite{Jacksohn.2019} reveal that income affected adoptions of PV systems. Instead of income, Petrovich et al. \cite{Petrovich.2019} find the availability of capital to be a good predictor for PV adoption intention.

Financial effects on the future expenditures of the decision-makers such as the reduction of fuel costs through the replacement of a heating system \cite{Curtis.2018} and reduced operational costs with an EV \cite{Rezvani.2015} can encourage adoption, what is confirmed by Gaspar and Antunes \cite{Gaspar.2011} who state that \q{although people value the cost of an appliance, they also value the long-term savings that can be achieved due to the lower energy consumption of [...] equipment}. On the other hand, negative long-term effects can impede adoption as shown by Kowalska-Pyzalska \cite{KowalskaPyzalska.2018} who find that higher prices for green electricity hinder a large share of decision-makers. The adoption of green electricity was instead found to be positively correlated with income \cite{KowalskaPyzalska.2018b, KowalskaPyzalska.2018, DiazRainey.2011}. 

In the case of PV systems, the positive effect of future revenues was found to be substantially smaller than the negative effect of the initial investment \cite{Jacksohn.2019}. This is in line with concepts suggesting a dilemma situation between direct negative and delayed positive outcomes, causing seemingly irrational decisions \cite{Frederiks.2015, Jager.2006}. 85\% of survey respondents of Simpson and Clifton \cite{Simpson.2017} indicate that in order to understand the costs and benefits of PV systems, education is needed, suggesting that grasping the true financial effect of a PV system is rather complicated. The same appears to apply to EVs, as \q{few households understood the financial calculations behind payback of investments}, leading consumers to \q{underestimate gasoline costs and savings over time} \cite{Ameli.2015}. Also, Danielis et al. \cite{Danielis.2018} and Rezvani et al. \cite{Rezvani.2015} suggest that decision-makers generally lack the capability to calculate \acrfull{TCO} of EVs. Even for energy efficient appliances, Mills and Schleich \cite{Mills.2012} and Baldini et al. \cite{Baldini.2018} suggest providing apps or features to \q{convey a simplified trade-off between energy efficiency and cost, such as payback times on a price-premium} in addition to labelling metrics.

\subsection{Environmental considerations}
\label{S:3.2}
Before purchasing an LCT, consumers presumably consider the expected externalities of adoption. Contributing to a better environment or the mitigation of climate change is a central reason for replacing heating systems \cite{Curtis.2018}, and adopting PV systems \cite{Jager.2006, Faiers.2006, Palm.2018b}. However, assessing the true environmental impact of an LCT is a difficult endeavour, and no transparent and reliable information on this matter is typically available for consumers, potentially leading to different evaluations of the environmental effect of single LCTs. For example, whilst some consumers are driven by the idea of protecting the environment when adopting an EV, others \q{expressed doubt about positive environmental consequences of EV adoption} due to battery production and electricity generation \cite{Rezvani.2015}.

Typically, no holistic life-cycle assessment of LCTs is publicly available, making it impossible to evaluate LCTs according to their true environmental impact. As highlighted by Metta et al. \cite{Metta.2020}, a technology can then be considered “low-carbon”, if it emits less carbon than conventional solutions; the differentiation is made by means of fossil-fueled solutions serving as a benchmark. Yet, whilst multiple products and services qualify as LCTs when compared to benchmark technologies, the effective improvement of residential GHG emissions depends on the previous state of the decision subjects home and their practices, and the rebound effect could reduce the expected reduction, or even cause a backfiring effect \cite{Aydin.2017}. For simplicity of the decision framework, we rely on the idea that all LCTs have the potential to reduce GHG emissions.

Instead of asking respondents about their perception of the environmental externalities of an LCT, the decision-makers attitude towards environmental issues is typically used as an antecedent of adoption decisions, relating to VBN theory.

A positive relationship between general environmental concerns and attitudes and the propensity to adopt green electricity has been found oftentimes, see e.g. \cite{KowalskaPyzalska.2018b, KowalskaPyzalska.2018, DiazRainey.2011}. Oppositely, Ozaki \cite{Ozaki.2011} find no relation between general environmental concerns and intention to adopt green electricity, whereas the willingness to accept cuts in living standard and higher prices to protect the environment are positively correlated with intention to adopt. Jansson \cite{Jansson.2011} find that adopters of alternative fuel vehicles exhibit higher personal moral norm (to act to protect the environment) than non-adopters, and according to Best et al. \cite{Best.2019}, environmental preferences are positively associated with intention and adoption of PV systems. Yet, although large proportions of the public express concern for the environment, e.g. in Australia \cite{Hobman.2014} and Germany \cite{BMU.2020}, attitude-consistent behavior is seldom observed. Thus, previous pro-environmental behavior has been used as a proxy to better grasp the decision-makers' willingness to comply with stated values.

Environmental behaviors have been found to be positively associated with PV adoption \cite{Best.2019}, experience in investing in green energy is positively associated with WTP for green energy and dynamic tariffs \cite{KowalskaPyzalska.2018b}, and Dieu-Hang et al. \cite{DieuHang.2017} find consumers who engaged more in energy saving practices and had higher knowledge of energy costs or efficiency labels to be more likely to adopt EEA. Contrarily, environmental behavior is not correlated with WTP for green electricity \cite{KowalskaPyzalska.2018}. Because pro-environmental behaviors are diverse, and energy saving behaviors could be motivated financially or by environmental considerations, we propose to separate environmental behaviors into environmentally driven behaviors (environmental behavior) and such behaviors that result in net monetary savings (energy expenditure sensitivity).

\subsection{Symbolic considerations}
\label{S:3.3}
According to Axsen and Kurani \cite{Axsen.2012}, consumers strive to express self-identity or convey status and associate symbolic benefits or barriers with LCTs.
Researchers revealed that the adoption of conspicuous LCTs such as the Toyota Prius \cite{Sexton.2014, Delgado.2015} and PV systems \cite{Dastrup.2012, Schelly.2014} is partly driven by the signaling of environmental consciousness. According to Hartmann und Apaolaza-Ib\'{a}\~{n}ez \cite{Hartmann.2012}, this holds not true for green electricity tariffs, as \q{status motives increase desire for green products only when consumed in public, but not in private}. Also, Arkesteijn und Oerlemans \cite{Arkesteijn.2005} argue that \q{the non-visibility of green electricity makes it impossible to gain social status by using green electricity}. For EVs, the perception of symbolic benefits and barriers vary strongly among decision-makers. Some relate EVs to a slow-moving lifestyle and feelings of embarrassment, others perceived a gain in social identity \q{related to a forward-thinking, modern and technology-oriented personality}, and some others don't want to be \q{associated with the green-driving identity} although they are concerned for the environment \cite{Rezvani.2015}.

\subsection{Normative considerations}
\label{S:3.4}
A number of psychological phenomena summarized under the term normative social influences affect consumers' consumption patterns \cite{Frederiks.2015} and residential LCT adoption (for a recent review of evidence see \cite{Wolske.2020}, for a review of social influence theories see \cite{Axsen.2012}).

There is strong evidence for the existence of spatial peer effects for PV systems resulting in regional clustering \cite{Rode.2020, Baginski.2019, Kosugi.2019, Graziano.2019} and partly for EVs \cite{Rezvani.2015}, referring to the concept of descriptive social norm that proposes behavior being driven by what is deemed normal \cite{Peattie.2010}. However, explaining regional clustering with peer effects based on spatial proximity \q{rules out any other explanation for why their behaviors might be correlated} \cite{Wolske.2020b}. Whilst Rode and Müller \cite{Rode.2020} conclude that \q{PV systems nearby may be seen as a large diverse pool of information which reduced uncertainty}, Baginski and Weber \cite{Baginski.2019} suggest that instead of being mainly driven by social imitation, spatial spillover is driven by unobserved regional characteristics such as the share of detached houses and electricity demand. A second type of social norm, the injunctive social norm which describes what decision-makers perceive to be expected and approved by peers \cite{Peattie.2010}, is found to positively affect WTP for and intention to adopt green electricity \cite{KowalskaPyzalska.2018, Ozaki.2011}, EVs \cite{Jansson.2011, Rezvani.2015} and PV systems \cite{Aggarwal.2019, Petrovich.2019}.

Normative considerations relate the behavior of the individual to its social context. DOI proposes that the relative timing of adoption in a social network is determined by the decision unit's ability to deal with risk and uncertainty and its venturesomeness - its innovativeness. The expression of the traits decreases from innovators to laggards \cite{Rogers.2003}. Additional adopters in decision units social context reduce uncertainty, and make an innovation gradually more attractive to decision units that are less like innovators and more like laggards. This is in line with \cite{Scheller.2021b, Scheller.2021}, stating that communication with peers plays an important role in PV adoption. Consequently, both, the decision units innovativeness, and the diffusion stage of an LCT within its social context matter, which was found particularly in the case of PV \cite{Rai.2016, Jager.2006}. Moreover, Scheller et al. \cite{Scheller.2020} reveal that the share of previous adopters in a decision-maker’s social network has an impact on how influential stakeholders like the social network, local utilities or consulting agencies are relative to each other, making it even more important to carefully elicit the local diffusion stage.

For innovativeness, Manning et al. \cite{Manning.1995} propose the two separate concepts of novelty seeking (the degree of being drawn to novel goods) and independent judgment making (the ability to make decisions independently of others). Jansson et al. \cite{Jansson.2011} find that adopters of alternative fuel vehicles exhibit higher novelty seeking than the general population, whereas no difference was found for independent judgment making. A positive effect of novelty seeking was found on the propensity to adopt \cite{Bashiri.2018} and interest in \cite{Wolske.2017} PV systems, and Petrovich et al. \cite{Petrovich.2019} find that likely adopters of PV systems tend to have more competence and interest in technical innovations than likely non-adopters, relating to the concept of novelty seeking.

\subsection{Behavioral effort considerations}
\label{S:3.5}
The issue of the compatibility of an LCT in the everyday lives of consumers and their habits, directly relating to DOI, has been found as an important contributing factor for potential adopters.
On the one hand, the initial effort required to adopt and put an LCT into use is found to negatively affect adoption, for example in the case of green electricity, where switching the provider is deemed difficult \cite{KowalskaPyzalska.2018b, Ozaki.2011}, PV adoption \cite{Korcaj.2015}, and micro-generation technologies in general \cite{Balcombe.2013}. Oppositely, the disruption caused by a replacement of a heating system was not found to be a relevant barrier \cite{Curtis.2018}. However, \q{from a behavioral perspective it is much easier to change a singular investment decision than to change daily behavior} \cite{Mills.2012}, shifting focus on required long-term behavioral changes related to the operation of LCTs. Negative effects of long-term behavioral efforts were found for EVs, where range limitations and charging require behavioral changes of consumers \cite{Rezvani.2015, Sierzchula.2014}. Coded reversely, the ease with which respondents believe to integrate PV systems in their lifestyle was found to positively affect adoption \cite{Aggarwal.2019}.

\subsection{Technical conditions}
\label{S:3.6}
Factors that received surprisingly low attention are the technical conditions required to adopt an LCT \cite{Geels.2017, Galvin.2020}. PV systems can be adopted only by consumers with decision power over a rooftop, and roof orientation appears to cause \q{a degree of self-selection among would-be prosumers} \cite{Galvin.2020}. Additionally, latitude, rooftop inclination and shading affect the electricity yield which also determines the financial return of the investment. In this vein, Korcaj et al. \cite{Korcaj.2015} find \q{the most often reported barrier impeding PV adoption is lack of resources such as [...] a suitable roof}. Similarly, the adoption of an EV requires the possibility to install a wall-box, which is not necessarily possible for renters, or consumers living in urban apartment complexes. Curtis et al. \cite{Curtis.2018} identify proximity to a networked fuel as key determinant of home-heating choice, and 7.5\% of non-adopters of green tariffs in Australia named limited availability as reason for not subscribing \cite{Hobman.2014}, all pointing towards the importance of the embedding of the decision object in the direct socio-technical system of the decision unit. Furthermore, multiple LCTs are newer, less carbon intensive versions of older technologies that deliver the same function and tend to be replaced most commonly when they breakdown, as has been found to be the case for EEA \cite{Wijaya.2013} and heating systems \cite{Curtis.2018}.

\subsection{The decision framework}
\label{S:3.7}

\begin{figure}[ht]
    \centering
    \includegraphics[width=0.75\textwidth]{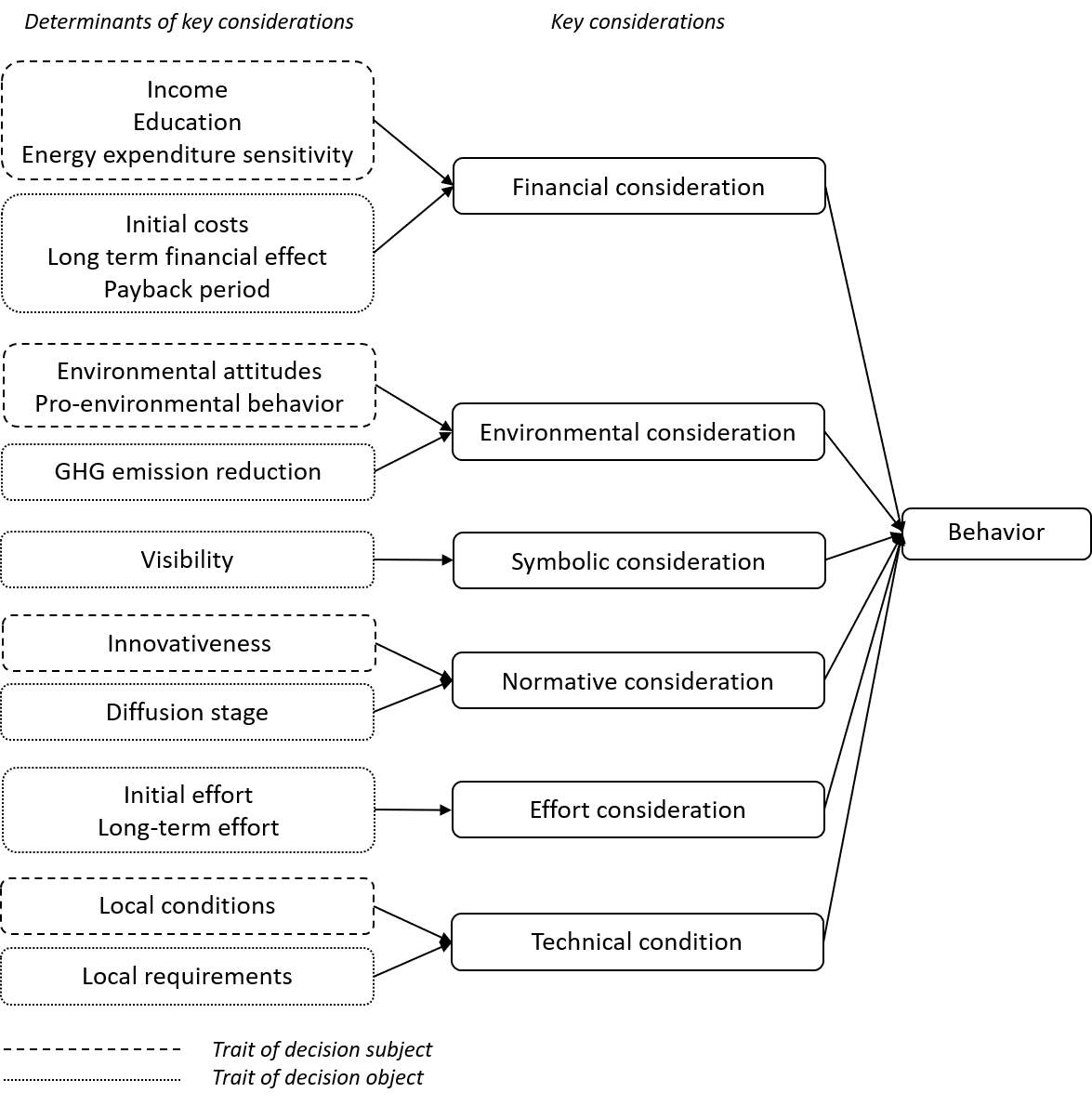}
    \caption{Derived decision framework for residential LCT adoption with 7 subject traits, 9 object traits and 6 considerations preceding adoption behavior.}
    \label{fig:ext_struc}
\end{figure}

Based on the above analysis, the general structure proposed in Figure \ref{fig:gen_struc} is developed into a decision framework with seven subject traits and nine object traits which is depicted in Figure \ref{fig:ext_struc}. The following six relations are suggested:

\textbf{R1:} Decision subjects with high energy expenditure sensitivity perceive a financial driver for LCTs with a short payback period. The higher initial costs, and the longer the payback period of an LCT, the more important income and education are to perceive a financial driver for adoption. A financial barrier is perceived for LCTs with a negative long-term financial effect, which applies more to decision subjects with lower income.

\textbf{R2:} The potential to reduce GHG emissions inherent in all LCTs, and the decision subjects environmental attitudes and previous pro-environmental behavior positively affect the perception of an environmental driver.

\textbf{R3:} A symbolic driver is perceived for visible LCTs.

\textbf{R4:} The higher the diffusion stage of an LCT, the stronger decision subjects with low innovativeness perceive a social norm.

\textbf{R5:} The higher initial and long-term effort of an LCT, the stronger perceived effort hinders behavior.

\textbf{R6:} Local requirements of an LCT must be met by the local conditions of a decision subject to avoid a technical barrier.

\section{Verification of the decision framework}
\label{S:4}

\subsection{The object - Evaluation of PV, EEA and GT concerning the proposed LCT traits}
\label{S:4.1}

A number of studies on PV mention the high initial costs \cite{Jager.2006, Bashiri.2018, Jacksohn.2019, Rai.2013}, expected revenues \cite{Jacksohn.2019} and long payback period \cite{Jager.2006, Jacksohn.2019}. Compared to alternative, non-efficient household appliances, EEAs typically require slightly higher initial investments, however, bring about energy savings in the near future, causing a positive financial effect after a relatively short payback period \cite{Gaspar.2011, Niemeyer.2010}. Oppositely, GT do not require initial financial investments, yet typically have a negative long-term financial effect \cite{Ek.2008, DiazRainey.2011}. The trait GHG emission reductions is evaluated similar for PV, EEA and GT, as due to lacking comprehensive life-cycle assessments, the LCTs are all assumed to reduce GHG emissions. Due to their placement on rooftop of houses, visibility of PV is evaluated high, whereas EEA and GT are invisible for peers \cite{Arkesteijn.2005}. The trait diffusion stage in social network is not evaluated, as innovations do not diffuse evenly in societies \cite{Geels.2018}, and the diffusion stage consequently depends on time and space. The initial efforts required to adopt PV and GT are considered high, because time and effort have to be invested to acquire information and prepare adoption \cite{Jager.2006, Rai.2013, Rai.2015b, Ozaki.2011}. In contrast, although the purchase of EEA also requires some amount of effort, choosing an energy-efficient appliance over a not particularly efficient one doesn't add a considerable amount of effort. For EEA and GT, no long-term behavioral effort is expected, whereas the picture is less clear for PV. Whilst Vasseur and Kemp \cite{Vasseur.2015b} suggest that \q{there is no effort connected to use}, Zhai and Williams \cite{Zhai.2012} and Anugwom \cite{Anugwom.2020} both state that prosumers must take responsibility for operation, maintenance and technical issues. Furthermore, if feed-in tariffs are gained, administrative efforts arise. In contrast to PV, where homeownership and a suitable rooftop are central local requirements, there are no specific prerequisites necessary to adopt an EEA or a GT.

\begin{table}[htbp]
    \footnotesize
    \centering
  \caption{Evaluation of the exemplary LCTs concerning the selected LCT traits}
    \begin{tabular}{p{14em}p{5.39em}p{5.39em}p{5.39em}}
    \toprule
    \multicolumn{1}{r}{} & \textbf{PV} & \textbf{EEA} & \textbf{GT} \\
    \midrule
    Initial costs & High  & Medium & Low \\
    \midrule
    Long-term financial effect & Positive & Positive & Negative \\
    \midrule
    Payback period & Long  & Short & / \\
    \midrule
    GHG emission reduction & Yes  & Yes & Yes \\
    \midrule
    Visibility & High  & Low   & Low \\
    \midrule
    Diffusion stage in network & / & / & / \\
    \midrule
    Initial effort & High  & Low   & High \\
    \midrule
    Long-term effort & Medium & Low   & Low \\
    \midrule
    Local requirements & High  & Low   & Low \\
    \bottomrule
    \end{tabular}
  \label{tab:lct_evalu}
\end{table}

\subsection{Covered aspects of the decision framework - The use of explanatory variables related to traits of decision subject and key considerations}
\label{S:4.2}

\begin{table}[htbp]
    \footnotesize
    \centering
\begin{threeparttable}
  \caption{Presentation of the included studies with sample size, data source, and an overview over the inclusion of traits of decision subject and key considerations in the analysis.}
\begin{tabular}{clcccccccccccccccc}
    \toprule
          &       &       &       & \multicolumn{7}{c}{\textit{Traits of decision subject}} &       & \multicolumn{6}{c}{\textit{Key considerations}} \\
\cmidrule{5-11}\cmidrule{13-18}          & \textbf{Ref.} & \textbf{N} & \textbf{D} & \textbf{INC} & \textbf{EDU} & \textbf{INN} & \textbf{EA} & \textbf{EB} & \textbf{EES} & \textbf{LC} &       & \textbf{FC} & \textbf{ENC} & \textbf{SC} & \textbf{NC} & \textbf{EFC} & \textbf{TC} \\
    \midrule
    \multicolumn{1}{c}{\multirow{8}[16]{*}{\begin{sideways}\textbf{PV}\end{sideways}}} & \cite{Schelly.2020} & 1842  & S     & x     & x     &       & x     &       &       &       &       & x     & x     &       &       &       &  \\
\cmidrule{2-18}          & \cite{Zander.2020} & 333   & S     & x     & x     &       &       &       &       &       &       & x     & x     &       & x     &       &  \\
\cmidrule{2-18}          & \cite{GavaGastaldo.2019} & 114   & S     & x     &       &       & x     &       &       &       &       & x     & x     &       & x     & x     &  \\
\cmidrule{2-18}          & \cite{Jacksohn.2019} & 24136 & P     & x     & x     & x     & x     &       &       & x    &       & x     &       &       &       &       &  \\
\cmidrule{2-18}          & \cite{Kastner.2019} & 120   & S     & x     & x     &       &       &       &       &       &       & x     & x     &       & x     &       &  \\
\cmidrule{2-18}          & \cite{Bondio.2018} & 1009  & S     &       &       &       & x     &       &       &       &       & x     &       &       & x     &       &  \\
\cmidrule{2-18}          & \cite{Rai.2016} & 380   & S     & x     & x     &       &       &       &       &       &       & x     & x     & x     &       &       &  \\
\cmidrule{2-18}          & \cite{Jager.2006} & 197   & S     & x     & x     &       & x     &       &       &       &       & x     & x     &       & x     & x     &  \\
    \midrule
    \multirow{8}[16]{*}{\begin{sideways}\textbf{EEA}\end{sideways}} & \cite{Sen.2020} & 5686  & P     & x     & x     &       &       &       &       & x     &       &       &       &       &       &       &  \\
\cmidrule{2-18}          & \cite{Schleich.2019} & 15055 & S     & x     &       &       & x     &       & x     & x     &       &       &       &       &       &       &  \\
\cmidrule{2-18}          & \cite{Schleich.2019b} & 15055 & S     & x     & x     & x     & x     &       &       & x     &       &       &       &       & x     &       &  \\
\cmidrule{2-18}          & \cite{Baldini.2018} & 1716  & P     & x     &       &       &       & x     &       & x     &       &       &       &       &       &       &  \\
\cmidrule{2-18}          & \cite{Trotta.2018} & ?     & P     & x     & x     &       & x     & x     &       & x     &       &       &       &       &       &       &  \\
\cmidrule{2-18}          & \cite{DieuHang.2017} & 12202 & P     & x     & x     &       & x     &       & x     & x     &       &       &       &       &       &       &  \\
\cmidrule{2-18}          & \cite{Mills.2012} & 4915  & P     &       & x     &       &       &       &       &       &       &       &       &       &       &       &  \\
\cmidrule{2-18}          & \cite{Gaspar.2011} & 1432  & S     & x     & x     &       & x     & x     & x     &       &       & x     & x     &       &       &       &  \\
    \midrule
    \multirow{3}[6]{*}{\begin{sideways}\textbf{GT}\end{sideways}} & \cite{Ziegler.2020} & 2687  & S     & x     & x     & x     & x     &       &       & x     &       &       &       &       &       &       &  \\
\cmidrule{2-18}          & \cite{MacPherson.2013} & 9000  & P     & x     & x     & x     & x     & x     &       & x     &       &       &       &       &       &       &  \\
\cmidrule{2-18}          & \cite{Arkesteijn.2005} & 115   & S     & x     &       &       & x     & x     &       &       &       & x     & x     &       & x     & x     &  \\
    \bottomrule
\end{tabular}%
    
\begin{tablenotes}
    \scriptsize
        \item D: Data source; S: Survey; P: Panel
        \item INC: Income; EDU: Education; INN: Innovativeness; EA: Environmental attitude; EB: Environmental behavior; EES: Energy expenditure sensitivity; LC: Local conditions; FC: Financial consideration; ENC: Environmental consideration; SC: Status considerations; NC: Normative considerations; EFC: Effort considerations; TC: Technical considerations
        \item x: included
\end{tablenotes}
\end{threeparttable}
    \label{tab:frequency_obs}%
\end{table}%

The analyzed 19 studies vary considerably in size and range from samples of around 120 (e.g., \cite{Gaspar.2011, Kastner.2019, Arkesteijn.2005} to datasets including up to 24.000 respondents \cite{Jacksohn.2019}. Typically, studies relying on data from panels such as the German Socio-Economic Panel \cite{Jacksohn.2019}, the U.S. Energy Information Administration’s Residential Energy Consumption Survey \cite{Sen.2020} or the United Kingdom Understanding Society Survey \cite{MacPherson.2013} have larger samples than single studies. An exemption are the studies \cite{Schleich.2019} and \cite{Schleich.2019b} that both analyze aspects of a particularly large survey (N = 15.055) administered in eight EU countries.

Table \ref{tab:frequency_obs} presents sample size, data source, and the coverage of the derived traits of the decision subject and key considerations, indicated by an 'x'. It becomes apparent that traits of the decision subject are assessed more frequently than key considerations, with the prior being accounted for on average in 11, and key considerations in 5 out of 19 studies. Whereas traits of the decision subject are included in studies on all three LCTs, considerations relating to the evaluation of the decision object are almost exclusively accounted for in the realm of PV adoption. All eight PV studies assess financial motives, and all six PV surveys include environmental motives. Normative considerations have been assessed in five surveys and one panel study. Among the studies on EEA, \cite{Schleich.2019b} includes normative considerations, and \cite{Gaspar.2011} accounts for financial and environmental considerations. The remaining 6 studies focus on traits of the decision subject as antecedents of behavior. Considerations related to behavioral control (effort and technical) and symbolic meanings of adoption are rarely accounted for.

Among traits of decision subjects and overall, the most frequently considered variables are income (17/19), education (14/19) and environmental awareness (13/19). Of key considerations, financial (10/19), environmental (8/19) and normative (7/19) considerations are most prominently assessed. The least often used variables are technical considerations (0/19), symbolic considerations being accounted for in one PV study (1/19), effort related considerations (3/19), and the traits innovativeness, appearing twice among GT studies, and one time each in PV and EEA studies (4/19) and energy expenditure sensitivity (4/19). With one exemption, the respondents living conditions are assessed only in studies on EEA and GT.

\subsection{Summary of findings - Effects of explanatory variables related to traits of decision subject and key considerations}
\label{S:4.3}
In Table \ref{tab:key_cons} and Table \ref{tab:indi_chara}, an overview of the explicit variable names in the original studies, and the assigned values to summarize the findings for the covered aspects are presented.

The included studies on PV adoption show that PV adopters can be characterized as high-income, highly educated, environmentally concerned people. Whereas Zander \cite{Zander.2020} do not find a relation between income and adoption, the samples in \cite{Rai.2016, Jager.2006, Schelly.2020, Kastner.2019} and \cite{GavaGastaldo.2019} have higher incomes than the general population. Additionally, the samples of \cite{Rai.2016, Jager.2006, Zander.2020, Schelly.2020, Kastner.2019} are highly educated, and show high environmental awareness \cite{Schelly.2020, GavaGastaldo.2019, Bondio.2018}. The regression analysis of Jacksohn et al. \cite{Jacksohn.2019} reveals a significant positive effect of income, education and environmental concern on adoption. Also Jager \cite{Jager.2006} find a significant difference between adopters and non-adopters concerning problem awareness. According to Jacksohn et al. \cite{Jacksohn.2019}, being a houseowner and living in a rural environment significantly increase the odds of being a PV owner, whereas the big five personality trait openness has no significant effect.

For EEA adopters, results are less clear. Five authors find a significant positive effect of income on adoption \cite{DieuHang.2017,Sen.2020, Schleich.2019, Schleich.2019b, Baldini.2018} whereas two find no effect \cite{Trotta.2018, Gaspar.2011}. \cite{Sen.2020, Schleich.2019b, Mills.2012} reveal a significant positive effect of education on adoption, \cite{Trotta.2018, DieuHang.2017} find no effect, and \cite{Gaspar.2011} even reveal a significant negative effect of higher education level on EEA adoption. According to Schleich et al. \cite{Schleich.2019b}, risk aversion is unrelated to EEA adoption. \cite{Trotta.2018, DieuHang.2017, Schleich.2019, Schleich.2019b} find a significant positive effect of environmental attitudes on EEA adoption, whereas Gaspar and Antunes \cite{Gaspar.2011} reveal a significant negative relationship between environmental attitudes and EEA adoption. Environmental behavior in the form of volunteering with a conversation group \cite{Trotta.2018}, general environmental behavior \cite{Gaspar.2011}, light score and energy efficient behavior index \cite{Baldini.2018} are significantly positive related with EEA adoption. Energy expenditure sensitivity, measured as awareness of efficiency labels, trust in efficiency labels, index of energy saving habits \cite{DieuHang.2017}, use of efficient light bulbs, turn off stand by modes, using rechargeable batteries \cite{Gaspar.2011}, and the importance of energy costs during decision-making \cite{Schleich.2019} is related significantly positive with adoption. A variety of variables addressing local conditions have been assessed in six of the seven panel studies. Three studies find home ownership to affect EEA adoption significantly positive \cite{DieuHang.2017, Sen.2020, Schleich.2019b}. Whilst \cite{Trotta.2018, DieuHang.2017} find no effect of residence type, \cite{Sen.2020, Schleich.2019, Baldini.2018} find residents of single family or detached houses to be more likely to adopt. Living in an urban environment \cite{DieuHang.2017} and planning to move \cite{Schleich.2019b} reduce the odds of adopting an EEA, whereas living in a townhouse showed no, and living in a farmhouse a significant positive effect \cite{Baldini.2018}.

According to MacPherson and Lange \cite{MacPherson.2013} and Ziegler \cite{Ziegler.2020}, a higher income, and not receiving a winter fuel payment significantly increases the odds of adopting GT , whereas Arkesteijn und Oerlemans \cite{Arkesteijn.2005} find no relation between income and adoption. Whilst MacPherson and Lange \cite{MacPherson.2013} reveal significant positive effects of education and innovativeness on adoption, Ziegler \cite{Ziegler.2020} find no effects. Environmental concern \cite{MacPherson.2013} and the New Environmental Paradigm scale \cite{Ziegler.2020} do not affect GT adoption, whereas the attitude towards the environment \cite{Arkesteijn.2005}, an ecological identity \cite{Ziegler.2020} and environmental behavior \cite{MacPherson.2013, Arkesteijn.2005} appear to increase the odds of adoption GT. Home ownership is not related to GT adoption \cite{MacPherson.2013, Ziegler.2020}.

Overall, the literature body shows that PV adopters typically have a high socio-economic status and are environmentally concerned. Oppositely, it doesn't provide sufficient insights into the relation between innovativeness, pro-environmental behavior, energy expenditure sensitivity and local conditions and adoption. For EEA and GT adopters, the socio-economic status is less clear, however, they are characterized by strong pro-environmental attitudes and behavior. EEA adopters additionally tend to be sensitive towards energy expenditures.\\

For PV adopters, financial and environmental motives were important drivers of adoption, as shown by the consistently positive effects in the studies. A positive relation between normative considerations, including discussions with other owners \cite{Rai.2016}, buying of PV by neighbors/acquaintances \cite{Jager.2006}, descriptive and injunctive social norm \cite{Kastner.2015}, neighborhood effect \cite{GavaGastaldo.2019} and friend influence \cite{Bondio.2018} and PV adoption could be shown. Furthermore, effort-related considerations affect decision-making, as technical and administrative support \cite{Jager.2006}, and perceived low effort \cite{GavaGastaldo.2019} were motivations for adoption.

In their survey on EEA adoption, Gaspar and Antunes \cite{Gaspar.2011} reveal that long-term savings and environmental issues were reasons why consumers decided for an energy-efficient appliance. Schleich et al. \cite{Schleich.2019b} find a significant positive effect of social norm on EEA adoption. For GT, Arkesteijn and Oerlemans \cite{Arkesteijn.2005} reveal that a favorable perception of price and effort related to GT drive adoption, whereas perceived environmental effects and communication in the interpersonal network were unrelated to adoption. Due to the scarcity of information, no meaningful conclusion about what considerations are central for EEA and GT adoption can be made.

\begin{landscape}

\begin{ThreePartTable}
    \tiny
    \renewcommand{\arraystretch}{1.0} 
    \setlength{\tabcolsep}{4pt}
\begin{longtable}{l p{0.191\textwidth} l p{0.191\textwidth} l p{0.191\textwidth} l p{0.191\textwidth} l p{0.191\textwidth} l p{0.05\textwidth} p{0.001\textwidth}}

    \caption{Explanatory variables and their effects clustered into the six key considerations.}\\
    \label{tab:key_cons}\\
    \toprule
    . & \textbf{Financial} &     & \textbf{Environmental} &     & \textbf{Status} &     & \textbf{Normative} &     & \textbf{Effort} &     & \textbf{Technical} &  \\
    \midrule
    \endfirsthead

    \caption*{Explanatory variables and their effects clustered into the six key considerations (continued).}\\
    \toprule
    . & \textbf{Financial} &     & \textbf{Environmental} &     & \textbf{Status} &     & \textbf{Normative} &     & \textbf{Effort} &     & \textbf{Technical} &  \\
    \midrule
    \endhead
    
    \bottomrule
    \endfoot
    
    \bottomrule
    \endlastfoot
    
    \multicolumn{13}{l}{\textbf{PV}} \\
    \midrule
    \cite{Schelly.2020} & Expected energy bill reduction & $+$   & Positive impact on the environment & $+$ &     &     &     &     &     &     &     &  \\
        & Low or no upfront costs & +   &     &     &     &     &     &     &     &     &     &  \\
    \midrule
    \cite{Zander.2020} & Economic motivations & $+$   & Environmental motivations & $+$ &     &     & Peer effects & $0$   &     &     &     &  \\
    \midrule
    \cite{GavaGastaldo.2019} & Investment opportunity & $+$   & Environmental protection & $+$ &     &     & Neighborhood effect & $+$ & Low effort & $+$   &     &  \\
    \midrule
    \cite{Jacksohn.2019} & Costs & $++$  &     &     &     &     &     &     &     &     &     &  \\
        & Revenue & $++$  &     &     &     &     &     &     &     &     &     &  \\
    \midrule
    \cite{Kastner.2019} & Economic motives & $+$   & Ecological motives & $+$ &     &     & Descriptive norm & $+$ &     &     &     &  \\
        &     &     &     &     &     &     & Injunctive norm & $+$ &     &     &     &  \\
    \midrule
    \cite{Bondio.2018} & Concern over future electricity prices & $+$   &     &     &     &     & Friend influence & $+$ &     &     &     &  \\
        & Reduce bills & $+$   &     &     &     &     &     &     &     &     &     &  \\
        & Good investment & $+$   &     &     &     &     &     &     &     &     &     &  \\
        & Increased home value & $+$   &     &     &     &     &     &     &     &     &     &  \\
    \midrule
    \cite{Rai.2016} & Financial evaluation of the investment & $+$   & Reducing impact on the environment & $+$ & Staying at the frontier of technology & $+$ &     &     &     &     &     &  \\
        & Hedge against electricity rate increases & $+$   &     &     &     &     &     &     &     &     &     &  \\
    \midrule
    \cite{Jager.2006} & The grant on offer & $+$   & Contribution to a better natural environment & $+$ &     &     & Discussion with other owners & $+$ & Central organisation of grant request & $+$   &     &  \\
        & The increased value of my home & $+$   &     &     &     &     & Buying of PV by neighbours /acquaintances & $+$ & Technical support offered by municipality & $+$   &     &  \\
    \midrule
    \multicolumn{13}{l}{\textbf{EEA}} \\
    \midrule
    \cite{Sen.2020} &     &     &     &     &     &     &     &     &     &     &     &  \\
    \midrule
    \cite{Schleich.2019} &     &     &     &     &     &     &     &     &     &     &     &  \\
    \midrule
    \cite{Schleich.2019b} &     &     &     &     &     &     & Social norm & $++$ &     &     &     &  \\
    \midrule
    \cite{Baldini.2018} &     &     &     &     &     &     &     &     &     &     &     &  \\
    \midrule
    \cite{Trotta.2018} &     &     &     &     &     &     &     &     &     &     &     &  \\
    \midrule
    \cite{DieuHang.2017} &     &     &     &     &     &     &     &     &     &     &     &  \\
    \midrule
    \cite{Mills.2012} &     &     &     &     &     &     &     &     &     &     &     &  \\
    \midrule
    \cite{Gaspar.2011} & Long-term savings & $+$   & Environmental issues & $+$ &     &     &     &     &     &     &     &  \\
    \midrule
    \multicolumn{13}{l}{\textbf{GT}} \\
    \midrule
    \cite{Ziegler.2020} &     &     &     &     &     &     &     &     &     &     &     &  \\
    \midrule
    \cite{MacPherson.2013} &     &     &     &     &     &     &     &     &     &     &     &  \\
    \midrule
    \cite{Arkesteijn.2005} & Perception of price green electricity & $++$  & Perception of relative advantage & $0$   &     &     & Perceived importance of used communication network: interpersonal & $0$   & Perception of ease of switching and use & $++$  &     &  \\
    
\end{longtable}

\begin{tablenotes}
\item ++: significant positive relation; +: positive relation; 0: no relation; -: negative relation; -- significant negative relation
\end{tablenotes}

\end{ThreePartTable}

\newpage

\begin{ThreePartTable}
    \tiny
    \renewcommand{\arraystretch}{1.0} 
    \setlength{\tabcolsep}{4pt}
\begin{longtable}{l p{0.13\textwidth} l p{0.13\textwidth} l p{0.13\textwidth} l p{0.13\textwidth} l p{0.13\textwidth} l p{0.13\textwidth} l p{0.13\textwidth} l}

    \caption{Explanatory variables and their effects clustered into the seven traits of the decision subject.}\\
    \label{tab:indi_chara}\\
    \toprule
    & \textbf{INC} &     & \textbf{EDU} &     & \textbf{INN} &     & \textbf{EA} &     & \textbf{EB} &     & \textbf{EES} &     & \textbf{LS} &  \\
    \midrule
    \endfirsthead

    \caption*{Explanatory variables and their effects clustered into the seven traits of the decision subject (continued).}\\
    \toprule
    & \textbf{INC} &     & \textbf{EDU} &     & \textbf{INN} &     & \textbf{EA} &     & \textbf{EB} &     & \textbf{EES} &     & \textbf{LS} &  \\
    \midrule
    \endhead
    
    \bottomrule
    \endfoot
    
    \bottomrule
    \endlastfoot

    \multicolumn{15}{l}{\textbf{PV}} \\
    \midrule
    \cite{Schelly.2020} & Income & + & Education & + &     &     & Climate change beliefs & +   &     &     &     &     &     &  \\
        &     &     &     &     &     &     &     &     &     &     &     &     &     &  \\
    \midrule
    \cite{Zander.2020} & Income & 0   & University degree & + &     &     &     &     &     &     &     &     &     &  \\
    \midrule
    \cite{GavaGastaldo.2019} & Income & + &     &     &     &     & Ecocentrism & +   &     &     &     &     &     &  \\
    \midrule
    \cite{Jacksohn.2019} & Income & ++ & Education level & ++ & Openness & 0   & Environmental concern & ++  &     &     &     &     & Houseowner & ++ \\
        &     &     &     &     &     &     &     &     &     &     &     &     & Rural & ++ \\
    \midrule
    \cite{Kastner.2019} & Income & + & Education level & + &     &     &     &     &     &     &     &     &     &  \\
    \cite{Bondio.2018} &     &     &     &     &     &     & Environmental concern & +   &     &     &     &     &     &  \\
    \midrule
    \cite{Rai.2016} & Income & + & Education & + &     &     &     &     &     &     &     &     &     &  \\
    \midrule
    \cite{Jager.2006} & Income & + & Education & + &     &     & Problem awareness & ++  &     &     &     &     &     &  \\
    \midrule
    \multicolumn{15}{l}{\textbf{EEA}} \\
    \midrule
    \cite{Sen.2020} & Income & ++ & Education & ++ &     &     &     &     &     &     &     &     & Ownership & ++ \\
        &     &     &     &     &     &     &     &     &     &     &     &     & Single family house & ++ \\
    \midrule
    \cite{Schleich.2019} & Income & ++ &     &     &     &     & Environmental identity & ++  &     &     & Importance of energy costs when deciding & ++ & Detached house & ++ \\
    \midrule
    \cite{Schleich.2019b} & Income & ++ & Education & ++ & Risk aversion & 0   & Environmental identity & ++  &     &     &     &     & Ownership & ++ \\
        &     &     &     &     &     &     &     &     &     &     &     &     & Likely move & -- \\
    \midrule
    \cite{Baldini.2018} & Income & ++ &     &     &     &     &     &     & Light score & ++  &     &     & Farmhouse & ++ \\
        &     &     &     &     &     &     &     &     & Energy efficient behavior index & ++  &     &     & Single family house & ++ \\
        &     &     &     &     &     &     &     &     &     &     &     &     & Townhouse & 0 \\
    \midrule
    \cite{Trotta.2018} & Income & 0   & Education & 0   &     &     & Environmental social norms & ++  & Volunteering with conservation group & ++  &     &     & Single family house & 0 \\
        &     &     &     &     &     &     & Knowledge of CO2 emissions & ++  &     &     &     &     &     &  \\
    \midrule
    \cite{DieuHang.2017} & Income & ++ & Education & 0   &     &     & Environmental concerns & ++  &     &     & Awareness of efficiency labels & ++ & Ownership & ++ \\
        &     &     &     &     &     &     & Support an environment group & ++  &     &     & Trust in efficiency labels & ++ & Urban environment & -- \\
        &     &     &     &     &     &     & Higher rank of environmentl issues & ++  &     &     & Index of energy-saving habits & ++ &     &  \\
    \midrule
    \cite{Mills.2012} &     &     & Completed high-school & ++ &     &     &     &     &     &     &     &     &     &  \\
        &     &     & Completed vocational school & ++ &     &     &     &     &     &     &     &     &     &  \\
        &     &     & Completed university & ++ &     &     &     &     &     &     &     &     &     &  \\
    \midrule
    \cite{Gaspar.2011} & Income & 0   & Education & -- &     &     & General environmental attitude & --  & General environmental behavior & ++  & Efficient light bulbs, turn off standby modes, rechargeable batteries & ++ &     &  \\
        &     &     &     &     &     &     &     &     & Fair trade, garbage separation, aprays/aerosols/chemicals avoidance, bags reuse, family-size packages use & 0 &     &     &     &  \\
    \midrule
    \multicolumn{15}{l}{\textbf{GT}} \\
    \midrule
    \cite{Ziegler.2020} & Higher household income & ++ & Education & 0   & Risk-taking preferences & 0   & New Environmental Paradigm & 0 &     &     &     &     & Ownership & 0 \\
        &     &     &     &     &     &     & Ecological identification & ++  &     &     &     &     &     &  \\
    \midrule
    \cite{MacPherson.2013} & Income & ++ & Education & ++ & Has renewable generation technology & ++ & Environmental attitude & 0 & Environmental behavior & ++  &     &     & Ownership & 0 \\
        & Receive winter fuel payment & -- &     &     &     &     &     &     &     &     &     &     &     &  \\
    \midrule
    \cite{Arkesteijn.2005} & Income & 0   &     &     &     &     & Attitude towards the environment & ++  & Actual displayed environmental behavior & ++  &     &     &     &  \\

\end{longtable}

\begin{tablenotes}
\item [1] INC: Income; EDU: Education; INN: Innovativeness; EA: Environmental attitude; EB: Environmental behavior; EES: Energy expenditure sensitivity; LC: Local conditions
\item [2] ++: significant positive relation; +: positive relation; 0: no relation; -: negative relation; -- significant negative relation
\end{tablenotes}

\end{ThreePartTable}

\end{landscape}
\section{Discussion}
\label{S:5}
\subsection{Relations between subject, object and context}
\label{S:5.1}
The results reveal, that whereas studies on PV adoption address both, traits of the decision subject and key considerations, studies on EEA and GT predominantly analyze traits of the decision subject. Across all LCTs, the most often analyzed variables are income, education and environmental attitudes. This is an interesting finding for itself, as it is surprising that despite it has oftentimes been shown that individual characteristics \cite{Ajzen.1991, Rogers.2003, Peattie.2010}, and particularly socio-demographics \cite{Arts.2011} are no precise predictors of specific behaviors, these variables received a considerable amount of attention.

Based on the results of the evaluation of the example LCTs and the systematic literature analysis, we now discuss to what extent the proposed decision framework can be confirmed or rejected, employing the relations explicitly formulated in Section \ref{S:3.7}.

The results confirm the proposed relation between traits of the decision subject and object with financial considerations preceding adoption in \textbf{R1}. The studies on PV adoption show, that adoption was driven by financial considerations, and that PV adopters are typically highly educated and have a high income. This suggests that they belong to a group of individuals that (1) has the financial means to pay the high investment costs, and (2) is capable of overcoming the dilemma situation between direct negative and delayed positive outcomes caused by the long payback period. Oppositely, findings for adopters of the medium priced product EEA concerning income and education are less clear, however, they appear to be characterized as customers with a high sensitivity towards energy expenditures. Looking at the differences in LCT traits, this seems plausible, as EEA come with only marginally higher investment costs than conventional appliances, and the break-even period is significantly shorter, allowing also low-income customers to pay the higher price, and reducing the temporal discounting bias by shifting focus on the near long-term savings. Typical GT adopters have the financial means to adopt an LCT with negative financial long-term effects, suggesting that unfavorable financial considerations are overruled by other considerations.

Adopters of all three LCTs show strong environmental attitudes, and for PV adopters, environmental considerations such as \q{Reducing impact on the environment} \cite{Rai.2016} and \q{Ecological motives} \cite{Kastner.2019} motivated behavior. Thus, results confirm \textbf{R2}, suggesting that environmentally aware customers evaluate the environmental effect of adoption, resulting in a driving force.

Surprisingly, although LCT literature suggests that the adoption of visible LCTs is partly driven by status aspirations \cite{Dastrup.2012, Schelly.2014, Sexton.2014, Delgado.2015}, this aspect of the decision framework is accounted for only in \cite{Rai.2016}, where \q{Staying at the frontier of technology} was an important motive for PV adopters. Therefore, the analysis provides no conclusive evidence for the relation between the decision objects visibility and status considerations proposed in \textbf{R3}.

Normative considerations such as injunctive and descriptive social norm \cite{Kastner.2019} and \q{Discussion with other owners} \cite{Rai.2016} were found to be positively related with PV adoption. Results are scarce for EEA and GT. Because the locally specific diffusion stage in the network of respondents has not been addressed, proposed relations between diffusion stage, perceived norm and innovativeness in \textbf{R4} can neither be confirmed nor rejected.

The analysis further revealed, that perceived behavioral control is not typically accounted for in behavioral studies, potentially because adopters have already overcome potential barriers. Yet, PV adopters report that administrative and technical support motivated adoption \cite{Jager.2006}, suggesting that a proactive reduction of the LCT trait initial effort, e.g. through support, can reduce the barrier related to effort. Furthermore, adopters appear to perceive effort related to PV adoption as not particularly high \cite{GavaGastaldo.2019}. Similarly, GT adopters were found to rate the behavioral effort to switch as low \cite{Arkesteijn.2005}. These findings contradict the proposed relation between initial and long-term effort and effort considerations, as the initial effort to adopt PV and GT has been evaluated as high. \textbf{R5} is thus rejected. It could cautiously be hypothesized that a trait of the adopters caused them to perceive the effort as low, whereas non-adopters perceive a barrier to adoption.

Despite a match of local requirements and local conditions are central considerations preceding PV adoption, where local requirements are high, this aspect of the decision framework was not accounted for in PV studies. Contrarily, the majority of EEA and GT studies investigated local conditions, such as homeownership, residence type and region, even though the adoption of EEA and GT doesn't require any particular local configurations. Therefore, \textbf{R6} can neither be confirmed nor rejected.

\subsection{Shortcomings of the methodology}
\label{S:5.2}
One main shortcoming of the present paper is the lack of an empirical component. Yet, the underlying questions we aim to address by the means of the decision framework can hardly be answered within one empirical survey, and the heterogeneity of the literature body hampered a statistical meta-analysis, as has also been pointed out by Kastner and Stern \cite{Kastner.2015}. Therefore, using existing literature on different LCTs appeared to be a suitable method to initially examine the suitability of the proposed decision framework.
Another shortcoming of the analysis is the small literature base in the main analysis (Sections \ref{S:4.2} and \ref{S:4.3}), which resulted from the decision to only integrate studies on actual behavior. This decision had been made contrary to approaches of e.g. \cite{Selvakkumaran.2019, Alipour.2020} because predictors vary among stages of decision-making, and intention is not necessarily translated into behavior \cite{Arts.2011, Peattie.2010, Carrington.2014}.

\subsection{Pledge for a more coordinated and integrated approach in research on  adoption}
\label{S:5.3}
In order to reach ambitious CO2 reduction goals, political, economic, societal, and scientific interest in the residential uptake of LCTs across the world will not diminish. Instead, the numbers of studies performed to understand who adopts which LCT for what reasons will increase, and deliver insights into the determinants of the adoption of for example EVs, PVs, home insulation measures, heating systems, LED light bulbs and efficient appliances. However, taking single snapshot of certain aspects of the respondents complex reality doesn't enhance understanding of residential LCT adoption, but rather increases the already existing \q{well-informed confusion} \cite{Peattie.2010}, as decision subject, object, and context will change over time, reducing the significance of the single studies results. Therefore, we pledge for a more coordinated and integrated approach in research on LCT adoption by shifting focus to the underlying relations between decision subject, decision object and context. 

Our analysis shows that the general structure of the proposed decision framework (Figure \ref{fig:gen_struc}), suggesting that traits of the decision object and subject are determinants of financial, environmental, symbolic, normative, effort and technical considerations, is helpful to analyze residential decision-making concerning LCTs. However, the analyzed literature revealed a number of blind spots concerning both, the role of and trade-offs between considerations for adoption, and the role of traits of decision subject and object for the outcome of considerations, hampering conclusive findings about structural differences in adoption decisions across e.g. high- and low-cost, or visible and invisible LCTs.
Therefore, we suggest to coordinate future studies by consistently including the identified traits of the decision subject, and six key considerations. Furthermore, traits of the decision object should be evaluated carefully to embed the decision in the local socio-technical context. This more coordinated approach could lead to a greater understanding of the underlying relations between subject, object and context. Central questions that remain to be answered are:

\begin{enumerate}
    \item To what extent are adoption decisions driven or hindered by the six key considerations?
    \item How do initial costs, long-term financial effects and payback period of decision objects and income, education and energy expenditure sensitivity affect the perception of a financial driver or barrier?
    \item Does the visibility of the decision object affect the perception of symbolic benefits?
    \item Do perceived norms depend on the local diffusion stage, and do innovative decision subjects adopt in earlier diffusion stages?
    \item Are the initial and the long-term effort related to adoption affecting the perceived effort, and what traits of the decision subject interfere?
    \item How important is a match between local requirements and local conditions for technical considerations?
\end{enumerate}

Moreover, we suggest to integrate findings of studies on the variety of LCTs in meta-analyses, building upon the work of Kastner and Stern \cite{Kastner.2015}, Bamberg and Möser \cite{Bamberg.2007} and Klöckner \cite{Klockner.2013}. In the meta-analyses, proposed traits of the decision object can be employed as moderators to investigate underlying relations on a higher level, confirming or rejecting the hypothesis that drivers and barriers for or against the adoption of an LCT are similar for LCTs with similar traits, and different for LCTs with different traits.

\section{Conclusion}
\label{S:6}
This research paper is an initial effort towards a fundamental understanding of relations between the decision subject (= who decides), decision object (= what is decided upon) and context (= when and where it is decided) in the context of residential LCT adoption. To this end, we took the perspective that the decision subject and object are equal parts of residential adoption behavior, and that a subjective evaluation of the object precedes adoption. The proposed decision framework suggests that considerations related to financial, environmental, symbolic, normative, effort and technical issues precede adoption, and that the outcome and importance of the consideration are determined by a set of traits of decision subject and object. Of the six proposed relations, two could be confirmed (financial and environmental), one could be rejected (effort), and three could neither be confirmed nor rejected due to lacking evidence.

The research provides fundamental practical implications. First, policies can enhance drivers and/or reduce barriers by targeting decision subject or object. Second, the six considerations emphasize that neither financial instruments, nor enhancing pro-environmental attitudes naturally translate into increased adoption rates, as other barriers might be present, and additional drivers might lack. Third, not one consumer group is most likely to adopt LCTs, but the interplay of decision subject and object determines who is most likely to adopt which product at what point in time.

The analysis has been performed as an intermediate step for the development and parametrization of an agent-based model (cf. \citep{johanning2020modular, Scheller.2019, reichelt2021towards}) that aims to simulate residential LCT adoption decisions in specific spatial-temporal contexts.

\section*{Declarations of interest} 
There are no competing interests.

\section*{Acknowledgement}
Emily Schulte and Fabian Scheller receive funding from the project SUSIC (Smart Utilities and Sustainable Infrastructure Change) with the project number 100378087. The project is financed by the Saxon State government out of the State budget approved by the Saxon State Parliament. Fabian Scheller also kindly acknowledges the financial support support of the European Union's Horizon 2020 research and innovation programme under the Marie Sklodowska-Curie grant agreement no. 713683 (COFUNDfellowsDTU).




\newpage

\bibliographystyle{elsarticle-num}
\bibliography{ModelDescription.bib}

\end{document}